\documentclass[a4paper,12pt]{revtex4}
\usepackage[latin1]{inputenc}
\usepackage{amsmath}
\usepackage{graphicx} 
\usepackage{subfigure} 
\usepackage{amsfonts}
\usepackage{hyperref} 
\usepackage{verbatim} 

\newtheorem{prop}{PROPOSITION}[section]

\begin{document}
\title{Mutually unbiased triplets from non-affine families of complex Hadamard matrices in dimension six}
\author{D. Goyeneche}
\email{dgoyeneche@cefop.udec.cl}
\affiliation{Departamento de Fis\'{i}ca, Universidad de Concepci\'{o}n, Casilla 160-C, Concepci\'{o}n, Chile\\Center for Optics and Photonics, Universidad de Concepci\'{o}n, Casilla 4016, Concepci\'{o}n, Chile}\vspace{1cm}

\begin{abstract}
We study the problem of constructing mutually unbiased bases in dimension six. This approach is based on an efficient numerical method designed to find solutions to the quantum state reconstruction problem in finite dimensions. Our technique suggests the existence of previously unknown symmetries in Karlsson's non-affine family $K_6^{(2)}$ which we confirm analytically. Also, we obtain strong evidence that no more than three mutually unbiased bases can be constructed from pairs which contain members of some non-affine families of complex Hadamard matrices.
\end{abstract}
\maketitle
Keywords: Mutually unbiased bases, Complex Hadamard matrices, Non-affine families.

\section{Introduction}
The existence of maximal sets of mutually unbiased (MU) bases in every dimension is a very important open problem in foundations of quantum mechanics. Two orthogonal bases are MU if they are as different as possible in Hilbert space, in the sense that the projection of every element of the first base onto every element of the second one has the same absolute value. This kind of bases has several applications in quantum information theory: quantum key distribution protocols \cite {Bennett,Brub,Cerf}, entanglement detection \cite{Spengler}, dense coding, teleportation, entanglement swapping, covariant cloning and state tomography (see \cite{Durt} and references therein). They are also interesting in mathematics since their connection with affine planes \cite{Gibbons} and finite geometries \cite{Bengtsson3}. Additionally, they are useful to solve the Mean King Problem \cite{Aharonov}. In a Hilbert space of dimension $d$ we can construct maximal sets of $d+1$ MU bases when $d$ is prime or prime power. Otherwise, analytical \cite{Zauner,Archer,Jaming,Grassl} and numerical
\cite{Butterley,Bengtsson1,Brierley,Brierley3,Jaming2} efforts to construct $d+1$ MU bases fail and it is suspected that they do not exist. The lowest dimension where this problem remains open is six, where most of the previously mentioned works have tried to find a solution. This paper presents a new method that numerically solves the problem to find the maximal set of MU bases that can be obtained from a given pair of MU bases. The most important advantage of our method is that the computational cost is independent of the pair of MU bases considered. Our method is not an algorithm because it does not stop with a definite answer but it converges very quickly even in higher dimensions.

This work is organized as follows: In Section II, we briefly introduce complex Hadamard matrices and mutually unbiased bases. In Section III, we present the method to find MU vectors and we discuss its convergence. We successfully test our approach in Section IV by searching the known 48 vectors MU to the identity and the Fourier matrix in dimension six, and we obtain known triplets containing the identity and the Di\c{t}\u{a} matrix. Section V contains our main results: we could not find triplets of MU bases from considering complex Hadamard matrix belonging to the non-affine families $K_6^{(2)}$ and $K_6^{(3)}$ existing in dimension six as well as other families contained in them. This study have allowed us to find new symmetries for the family $K_6^{(2)}$.

\section{Complex Hadamard matrices and mutually unbiased bases}
This section contains the minimal information about complex Hadamard matrices and mutually unbiased bases required to make the paper self-contained; more details can be found in \cite{Bengtsson1} and \cite{Brierley4}, for example. Two orthonormal bases $\{|\varphi_k\rangle\}$ and $\{|\phi_l\rangle\}$ defined on a $d$-dimensional Hilbert space are mutually unbiased (MU) if they satisfy the property
\begin{equation}
|\langle\varphi_k|\phi_l\rangle|^2=\frac{1}{d},
\end{equation}
for every $k,l=0,\dots,d-1$. Maximal sets of $d+1$ MU bases have been found in every prime \cite{Ivanovic} and prime power \cite{Wootters} dimension. A lower bound on their members can be established in the general case of $d=p_1^{r_1}\dots p_n^{r_n}$, where  $p_1^{r_1}<\dots <p_n^{r_n}$: it is known how to construct at least $p_1^{r_1}+1$ MU bases \cite{Bengtsson2}. Here, $d=p_1^{r_1}\dots p_n^{r_n}$ is the prime power decomposition of the number $d$. In the particular case of $d=6$ the lower bound is three, and this is the maximal number of MU bases attained so far.

A square matrix is a complex Hadamard matrix if it has unimodular entries and orthonormal columns. Such matrices exist in every dimension and the Fourier matrices represent the simplest proof of their existence. For example, in dimension four the Fourier matrix is given by
\begin{equation}
F_4=
\left( \begin{array}{cccc}
1&1&1&1\\
1&\omega&-1&\omega^3\\
1&\omega^2&1&\omega^2\\
1&\omega^3&-1&\omega
 \end{array} \right),
\end{equation}
where $\omega=e^{2\pi i/4}$. Also, the tensor product of complex Hadamard matrices is a complex Hadamard matrix. The simplest example is given by the tensor product of two Fourier matrices defined in dimension two:
\begin{equation}
F_2\otimes F_2=\left( \begin{array}{rr}
1&1\\
1&-1
 \end{array} \right)\otimes\left( \begin{array}{rr}
1&1\\
1&-1
 \end{array} \right)=
\left( \begin{array}{rrrr}
1&1&1&1\\
1&-1&1&-1\\
1&1&-1&-1\\
1&-1&-1&1
 \end{array} \right),
\end{equation}
which gives us a real Hadamard matrix. Complex Hadamard matrices have been extensively studied in recent years and they are very hard to find when $d>5$ \cite{Brierley}. We say that two complex Hadamard matrices $H_1$ and $H_2$ are \emph{equivalent} ($H_1\sim H_2$) if there exist unitary diagonal operators $D_1,D_2$ and permutation operators $P_1,P_2$, such that
\begin{equation}
H_1=P_1D_1H_2D_2P_2.
\end{equation}
A complex Hadamard matrix may belong to a continuous set of inequivalent complex Hadamard matrices called \emph{a family}. A family of complex Hadamard matrices $H(x)$ is affine if it can be cast in the form
\begin{equation}\label{affine}
H(x)=H(0)\circ\mathrm{Exp}{(iR(x))},
\end{equation}
where $R(x)$ is a real matrix for all value of the real parameter $x=(x_1,\dots,x_s)$ and $\mathrm{Exp}$ is the entry-wise exponential function given by
\begin{equation}
    \mathrm{Exp}{(iR(x))}_{lm}=\exp{(iR(x)_{lm})},
\end{equation}
and the symbol $\circ$ denotes the Hadamard product $(A\circ B)_{lm}=A_{lm}B_{lm}$.
The number $s$ of independent parameters corresponds to the dimension of the family. If a continuous family of inequivalent complex Hadamard matrices is not affine we say it is \emph{non-affine}. For example, the families stemming from the Fourier matrix in dimension six $(F^{(2)}_6)$ and the Di\c{t}\u{a} family $(D^{(1)}_6)$ are affine families, whereas the Karlsson families $K^{(2)}_6$ and $K^{(3)}_6$ are non-affine. The notation here considered is consistent with the catalog of complex Hadamard matrices presented by Bruzda-Tadej-\.{Z}yczkowski \cite{Bruzda}. In this notation, the upper index denotes the dimension of the family and the lower index the dimension of the space where it is defined. If a complex Hadamard matrix $H$ belongs neither to an affine nor to a non-affine family we call it \emph{isolated}. In other words, it is impossible to obtain a complex Hadamard matrix inequivalent to $H$ from infinitesimal perturbations to its entries. A set of MU bases is \emph{inextensible} if no further orthonormal basis MU to every base of the set exists. A well-known fact is that any set of $d+1$ MU bases is inextensible, and it is conjectured that every triplet of MU bases in dimension six is inextensible. We mention that the complete set of inextensible MU bases in dimensions $d\leq5$ has been found in \cite{Brierley2} by considering Buchberger's algorithm \cite{Buchberger}. This algorithm is a generalization of gaussian elimination to non-linear multivariate polynomial equations. Also, a characterization of triplets of MU bases in $d=6$ have been given if the second MU basis belongs to an affine family of Hadamard matrices \cite{Brierley3}. In a recent paper, it has been analytically proven that given any triplet of MU \emph{product} bases in dimension six it is not possible to find even a \emph{single} vector MU to the triplet \cite{McNulty}.

From definition of MU bases, it is easy to show that any pair $\{\mathcal{B}_1,\mathcal{B}_2\}$ of MU bases is unitary equivalent to a pair $\{\mathbb{I},H\}$, where $\mathbb{I}$ is the identity matrix and $H$ is a complex Hadamard matrix. Therefore, the existence and classification of mutually unbiased bases is closely related to the existence of the maximal set of complex Hadamard matrices. Given a pair of MU bases $\{\mathbb{I},H\}$ the problem to find the complete set of vectors MU to both of them can, in principle, be solved by considering the Buchberger's algorithm. However, when $H$ belongs to a special kind of families of complex Hadamard matrices, known as \emph{non-affine}, even 16 GB were not enough memory for the algorithm to terminate and identify the solutions \cite{Brierley4}. In the next section we will present a method to find the complete set of MU vectors to a given pair of the form $\{\mathbb{I},H\}$ whose efficiency does not decrease for non-affine families.

\section{MU vectors as fixed points}\label{PIO}
\subsection{The physical imposition operator}
In this section, we present a method that allows the numerical construction of all of vectors MU to a given pair of MU bases $\{\mathbb{I},H\}$. The iterative method used here allows us to efficiently generate highly accurate approximations to the solutions of the defining set of equations. The desired states are attractive fixed points of the \emph{physical imposition operator} \cite{Goyeneche1} which has been used previously to find those quantum states known as Pauli partners \cite{Goyeneche2}. The problem has a unique answer if the given probability distributions are informationally complete; otherwise a finite or infinitely many number of solutions may exist. The physical imposition operator is useful for the problem studied here because the search for MU bases is a particular case of the \emph{quantum state reconstruction problem}, namely to determine the quantum state of a physical system compatible with probability distributions obtained from actual measurements.

To illustrate the concept of the imposition operator, let us assume that  $\{|\varphi_k\rangle\}$ and $\{|\phi_l\rangle\}$ are the eigenvectors bases of two observables acting in $\mathbb{C}^d$ say, $A$ and $B$, respectively.  We suppose that two probability distributions $\{p^{(A)}_k\}$ and $\{p^{(B)}_l\}$ have been obtained by measuring of the observables $A$ and $B$, respectively. For simplicity, the distributions are assumed to be given \emph{exactly}, which is only possible when the ensemble of quantum states is infinite. To reconstruct a pure state $|\Phi\rangle\in\mathbb{C}^d$ compatible with the measurements, we need to find all solutions $\{|\Phi\rangle\}$ of the following set of coupled non-linear equations:
\begin{eqnarray}
|\langle\varphi_k|\Phi\rangle|^2=p^{(A)}_k,\label{QSRP1}\\
|\langle\phi_l|\Phi\rangle|^2=p^{(B)}_l,\label{QSRP2}
\end{eqnarray}
where $k,l=0,\dots,d-1$. In order to find a solution we perform the following steps:
\begin{enumerate}
\item Choose a quantum state $|\Psi_0\rangle$ at random, which will be called the \emph{seed}.
\item Decompose the state $|\Psi_0\rangle$ in the basis $\{|\varphi_k\rangle\}$,
\begin{equation}
|\Psi_0\rangle=\sum_{k=0}^{d-1}c_k|\varphi_k\rangle.
\end{equation}
\item Modify the amplitudes of the expansion coefficients $c_k$ in order to impose the information about $A$,
\begin{equation}
c_k\rightarrow\sqrt{p^{(A)}_k}\frac{c_k}{|c_k|}.
\end{equation}
\end{enumerate}
In the last step, we have replaced the amplitudes of the coefficients $\{c_k\}$ compatible with those of the observable $A$; note that we did not modify the phase factors $c_k/|c_k|$ because we can not draw any conditions about them from the data $\{p^{(A)}_k\}$. The \emph{physical imposition operator} implements the transformations just described in one operation,
\begin{equation}
    T_{A,p^{(A)}}|\Psi_0\rangle=\sum_{k=0}^{d-1}
    \sqrt{p^{(A)}_k}\,\frac{\langle\varphi_k|\Psi_0\rangle}{|\langle\varphi_k|\Psi_0\rangle|}|\varphi_k\rangle;
\end{equation}
when $|\Psi_0\rangle$ happens to be orthogonal to the state $|\varphi_k\rangle$, for any $k=0,\dots,d-1$, we define
\begin{equation}\label{condition}
\langle\varphi_k|\Psi_0\rangle/|\langle\varphi_k|\Psi_0\rangle|\rightarrow1.
\end{equation}
This operator is non-linear and its action on every quantum state is well-defined. The action of this operator on a randomly chosen state $|\Psi_0\rangle$ can be interpreted as incorporating what we learn about the unknown state when the observable $A$ is measured. In other words, the initial state $|\Psi_0\rangle$ has no information about the quantum system considered while the state $T_{A,p^{(A)}}|\Psi_0\rangle$ contains all the information we have acquired by measuring $A$ in the unknown state. Note that $T_{A,p^{(A)}}$ is idempotent because applying it once exhausts the information available about $A$.

Next, we proceed in a similar way with the second observable $B$ in order to try to find a solution of the set of Eqs.(\ref{QSRP1},\ref{QSRP2}), defining the physical imposition operator associated with the observable B,
\begin{equation}
    T_{B,p^{(B)}}|\Psi_0\rangle=\sum_{r=0}^{d-1}
    \sqrt{p^{(B)}_r}\,\frac{\langle\phi_r|\Psi_0\rangle}{|\langle\phi_r|\Psi_0\rangle|}|\phi_r\rangle.
\end{equation}
Unfortunately, the state
\begin{equation}
     |\Psi_1\rangle=T_{B,p^{(B)}}(T_{A,p^{(A)}}|\Psi_0\rangle),
\end{equation}
generally does not contain the complete information about both $A$ and $B$: some of the information about $A$ is destroyed when $T_{B,p^{(B)}}$ is imposed, which is a consequence of the commutation rule $[A,B]\neq0$. If $A$ and $B$ commute, it is trivial to find a solution to Eqs. (\ref{QSRP1},\ref{QSRP2}), namely
\begin{eqnarray}
\mathcal{S}&=&\{|\Psi\rangle\in\mathcal{H}\,/\,|\Psi\rangle=T_{A,p^{(A)}}|\Psi_0\rangle,\,\forall\,|\Psi_0\rangle\in\mathcal{H}\}\\
&=&\{|\Psi\rangle\in\mathcal{H}\,/\,|\Psi\rangle=T_{B,p^{(B)}}|\Psi_0\rangle,\,\forall\,|\Psi_0\rangle\in\mathcal{H}\},
\end{eqnarray}
In general, the state $|\Psi_1\rangle$ has the complete information about $B$ and only partial information about $A$, so the composite operator $T_{B,p^{(B)}}T_{A,p^{(A)}}$ is not idempotent. Therefore, we can iterate the procedure just described and analyze the convergence of the sequence
\begin{equation}\label{SEQUENCE}
|\Psi_n\rangle=(T_{B,p^{(B)}}T_{A,p^{(A)}})^n|\Psi_0\rangle,\quad n \in\mathbb{N}.
\end{equation}
It has been proven \cite{Goyeneche2} that every solution of the system of coupled equations (\ref{QSRP1},\ref{QSRP2}) is an attractive fixed point of  $T_{B,p^{(B)}}T_{A,p^{(A)}}$. Moreover, this property also holds for a general set of observables $A,B,C,\dots$ and probability distributions $p^{(A)},p^{(B)},p^{(C)},\dots$ The iterations are robust under adding redundant information, and the sequences converge if and only if the probability distributions are compatible, in the sense that the Heisenberg uncertainty principle is not violated.

The problem of constructing MU vectors is a particular case of the quantum state reconstruction problem just described. Let $A$ and $B$ be two observables with a pair of MU eigenbases $\mathcal{B}_A$ and $\mathcal{B}_B$. A vector is MU to the pair $\{\mathcal{B}_A,\mathcal{B}_B\}$ if it has equally weighted probability distributions with respect to both observables, that is,
\begin{equation}
p^{(A)}_k=\frac{1}{d}\hspace{0.5cm}\mbox{and}\hspace{0.5cm}p^{(B)}_l=\frac{1}{d},
\end{equation}
for every $k,l=0,\dots,d-1$. Interestingly, when the eigenvectors bases are MU, every basin of attraction is found to be of the same size, verified numerically in every prime dimension $2\leq d\leq 37$ \cite{Goyeneche2}, as well as in every simulation reported below for $d=6$. This property indicates that the efficiency of the algorithm is maximal when the eigenvector bases of the observables are MU, because the number of randomly chosen seed states needed to find all solutions is minimized. This observation conforms with the idea that the redundancy of information is minimal when the observables have MU eigenvector bases.
\subsection{Convergence}\label{convergence}
In order to analyze the convergence of the sequence $|\Psi_n\rangle$ defined in Eq.(\ref{SEQUENCE}) we need to define a metric for quantum states. We want to determine when a solution given by our method is a solution of the coupled system of equations given by Eqs. (\ref{QSRP1},\ref{QSRP2}). Let $A$ be an observable having the eigenvectors base $\{|\varphi_k\rangle\}_{k=0,\dots,d-1}$ and let $|\phi\rangle$ and $|\psi\rangle$ be two arbitrary quantum states. The distance between the probability distributions associated with the observable $A$ in the states $|\phi\rangle$ and $|\psi\rangle$ can be defined by means of Hellinger's metric \cite{Hellinger},
\begin{equation}
\label{Hellinger}
D_A^2(|\phi\rangle,|\psi\rangle)=\sum_{k=0}^{d-1}\left(|\langle\varphi_k|\phi\rangle|-|\langle\varphi_k|\psi\rangle|\right)^2.
\end{equation}
This metric compares two probability distributions of the eigenvalues of a single observable and it is important to realize that this is a metric for probability distributions, \emph{not} for states. In the present context, we need to consider more than one observable and the corresponding probability distributions. Therefore, we introduce the Hellinger metric for $m$ observables, the so-called \emph{distributional metric} \cite{Goyeneche2},
\begin{equation}\label{distribucional}
\mathcal{D}^2_{A^1,\dots,A^m}(|\phi\rangle,|\psi\rangle)=\frac{1}{m}\sum_{j=1}^m D^2_{A^j}(|\phi\rangle,|\psi\rangle),
\end{equation}
where $D_{A^j}(|\phi\rangle,|\psi\rangle)$ is the Hellinger distance of the observable $A^j$, defined in Eq.(\ref{Hellinger}).

In our study of MU bases, we will always start from a pair of bases $\mathcal{B}_A=\{|\varphi_k\rangle,k=0,\dots,d-1\}$ and $\mathcal{B}_B=\{|\phi_l\rangle,l=0,\dots,d-1\}$. Now, assuming that $|\Phi\rangle\in\mathbb{C}^d$ is a vector MU to these bases, the expression
\begin{eqnarray}
\label{distri}
\mathcal{D}_{A,B}(|\Psi_n\rangle,|\Phi\rangle)&=&\sqrt{\frac{1}{2} D^2_{A}(|\Psi_n\rangle,|\Phi\rangle)+\frac{1}{2}D^2_{B}(|\Psi_n\rangle,|\Phi\rangle)}\nonumber\\
&=&\sqrt{\frac{1}{2}\sum_{k=0}^{d-1}\left(|\langle\varphi_k|\Psi_n\rangle|-\sqrt{\frac{1}{d}}\right)^2+\frac{1}{2}\sum_{l=0}^{d-1}\left(|\langle\phi_l|\Psi_n\rangle|-\sqrt{\frac{1}{d}}\right)^2}\nonumber\\
&=&\sqrt{2-\frac{1}{\sqrt{d}}\left(\sum_{k=0}^{d-1}|\langle\varphi_k|\Psi_n\rangle|+\sum_{l=0}^{d-1}|\langle\phi_l|\Psi_n\rangle|\right)},\label{distri}
\end{eqnarray}
tells us how close the state $|\Psi_n\rangle$ is to being MU to $\mathcal{B}_A$ and $\mathcal{B}_B$. We will say that a sequence has converged when
\begin{equation}\label{bound}
   \mathcal{D}_{A,B}(|\Psi_n\rangle,|\Phi\rangle)<0.01,
\end{equation}
which means that the absolute error of the amplitudes is less than $8\times10^{-4}$ on average. Numerical simulations suggest that the absolute error of every amplitude of a solution is very close to the averaged error just mentioned. Given that the desired solutions are (stable) attractive fixed points, our approximations must be close to the exact solutions of the problem. In the next section, we test our method by constructing known sets of states MU to a number of pairs consisting of the identity and a complex Hadamard matrix of order six.

\section{Testing the method: Tao, Fourier and Di\c{t}\u{a} matrices}\label{Fourier_section}

In this section, we apply the approach described above to four cases which have been studied before, reproducing successfully known results. We will (i) search for states simultaneously MU to the identity matrix $\mathbb{I}$ and Tao's matrix $S_6^{(0)}$, the only known isolated Hadamard matrix of order six; (ii) we will derive the complete set of vectors MU to the pair $\{\mathbb{I},F_6\}$ with the numerical results being, in fact, so accurate that we are able to deduce an interesting analytic result about this set; (iii) we are able to confirm that there are no quadruples containing members of the Fourier family $F_6(a,b)$, and (iv) we will search for states MU to the standard basis and members of the one-parameter Di\c{t}\u{a} family $D_6^{(1)}(c)$. The results presented in this section are summarized in the first three rows of Table \ref{tabla} in Sec. \ref{Summary}.

(i) \emph{Tao's matrix} $S_6$: The pair $\{\mathbb{I},S_6^{(0)}\}$ cannot be complemented by six orthogonal vectors to form a triplet of MU bases \cite{Brierley6}. We are able to confirm this result by unsuccessfully searching for a third basis by means of the imposition operator. We found 90 vectors MU to the pair $\{\mathbb{I},S_6^{(0)}\}$ but a third MU base cannot be constructed from them.

(ii) \emph{Fourier matrix} $F_6$: It is impossible to construct four MU bases which contain the pair $\{\mathbb{I},F_6\}$ \cite{Grassl,Brierley4,Bjorck1,Bjorck2}. More specifically, it is known that 48 vectors exist which are MU to this pair of bases, giving rise to 16 different ways to construct a triplet of MU bases.

We have been able to unambiguously identify 48 vectors MU to $\{\mathbb{I},{F_6}\}$, and they agree with the known solutions \cite{Bjorck1,Bjorck2,Grassl,Bengtsson1,Brierley6}. A careful analysis of the numerical expressions revealed that the components of 12 of the vectors can be expressed solely in terms of sixth roots of unity while the remaining 36 vectors also depend on Bj\"{o}rck's number,
\begin{equation}
	a=\frac{1-\sqrt{3}}{2}+i\sqrt{\frac{\sqrt{3}}{2}},
\end{equation}
which is unimodular, and occurs as $a,a^*,a^2,{(a^2)}^*$, where $*$ means complex conjugation. The analytic expression for the number $a$ has been found from the numerical results by imposing the unbiasedness of a solution to the pair $\{\mathbb{I},F_6\}$.

It turns out that the 48 vectors can be grouped into three sets, each corresponding to one orbit under the Weyl-Heisenberg group. To see this, let us first define the displacement operators $D_p$ by
\begin{equation}
    D_p=\tau^{p_1p_2}X^{p_1}Z^{p_2},
\end{equation}
with $p \equiv (p_1,p_2)\in\mathbb{Z}_d^2$, where $X$ and $Z$ are the \emph{shift} and \emph{phase} operators, respectively, defined by their actions on the states of the canonical basis,
\begin{equation}
    X|\varphi_k\rangle= |\varphi_{k+1}\rangle, \mbox{ and } Z | \varphi_k \rangle=\omega^k | \varphi_k \rangle .
\end{equation}

Three vectors generating the mentioned orbits under Weyl-Heisenberg group are given by
\begin{eqnarray}
v_1&=&\frac{1}{\sqrt{6}}(1, i, \omega^4, i, 1, i\omega^4),\\
v_2&=&\frac{1}{\sqrt{6}}(1,- i, \omega^2, -i, 1, -i\omega^2),\\
v_3&=&\frac{1}{\sqrt{6}}(1, i a, a^2, -i a^2, -a, -i),
\end{eqnarray}
with $\omega=e^{2\pi i/6}$; we noted that $v_1$ and $v_2$ are eigenvectors of $D_{(\mu,\mu)}$ and $D_{(\mu,5\mu)}$ respectively, for every $\mu=0,\dots,d-1$.
Consequently, the two orbits generated by $v_1$ and $v_2$ each have six elements only consisting of the so-called Gaussian states \cite{Bengtsson1} which, in fact, can be written as product vectors if we swap components 2 and 5 \cite{McNulty2}. We understand the origin of the Gaussian vectors very well: the eigenvectors of the operators $\{D_{(1,0)},D_{(0,1)},D_{(1,1)}\}$ -- that is, the bases consisting of the eigenvectors of $X$, $Z$ and $XZ$ -- form a triplet of MU bases in \emph{any} dimension \cite{Goyeneche3}. Every eigenvector of $XZ$ in dimension six is a member of the orbit generated by $v_1$, whereas the orbit generated by $v_2$ is its complex conjugated orbit, in agreement with the fact that $v_2=v_1^*$. Given a MU vector of the pair $\{\mathbb{I},F_6\}$ it is well known that its complex conjugate is also a MU vector in any finite dimension. Moreover, this is also valid in infinite dimension and it is related to Perelomov's conjecture about the existence of Pauli partners \cite{Moroz}.
The origin of the vector $v_3$, however, is not clear to us; it is not an eigenvector of any displacement operator, and it gives rise to an orbit with 36 different states.

Finally, in condensed form, the 48 MU vectors can be written as
\begin{eqnarray}
&\{D_{(\mu,0)}v_1\}_{\mu\in\mathbb{Z}_6}&\\
&\{D_{(\mu,0)}v_2)\}_{\mu\in\mathbb{Z}_6}&\\
&\{D_{(\mu,\mu\nu)}v_3\}_{\mu,\nu\in\mathbb{Z}_6}.&
\end{eqnarray}
Both the first and second set define a circulant matrix each while the last set give rise to six circulant matrices. These observations generalize to other dimensions $d$ as follows.
\begin{prop}\label{PropMU}
Let $| \phi \rangle $ be a state MU to the pair $\{\mathbb{I},F_d\}$, where $F_d$ is the Fourier matrix defined in dimension $d$. Then, the set $\{D_p | \phi\rangle \}_{p\in\mathbb{Z}_d^2}$ defines an orbit of MU vectors which has $d$ elements if $| \phi \rangle$ \emph{is} an eigenvector of any operator $D_p$ and it has $d^2$ elements if $| \phi \rangle$ \emph{is not} an eigenvector of any $D_p$.
\end{prop}

The proof of the proposition is trivial because the pair of bases of $\mathbb{C}^d$ defined by $\{\mathbb{I},F_d\}$ is invariant under the action of the displacement operators. Also, the eigenvectors of $D_{(\mu,\mu\nu)}$ are shifted cyclically under the action of $D_{(\mu',\mu'\nu')}$ for every $\mu'=0,\dots,d-1$ and $\nu'\neq\nu$ \cite{Bandyo}.

(iii) \emph{Fourier family} $F_6^{(2)}(a,b)$: we have attempted to extend pairs of the form $\{\mathbb{I},F_6^{(2)}(a,b)\}$ for $10^5$ randomly chosen values of $a$ and $b$, taken from the entire parameter range. In each case we found a triplet of MU bases and we could not find a \emph{single} additional vector MU to it.

(iv) \emph{Dita family} $D_6^{(1)}(c)$: The Di\c{t}\u{a} family $D_6^{(1)}(c)$ \cite{Dita} is an affine one-parameter family of complex Hadamard matrices in dimension six which is closely related to the Fourier family. Bengtsson \emph{et al.} \cite{Bengtsson1} found two triplets of MU bases which extend the pair $\{\mathbb{I},D_6^{(1)}(0)\}$. To do so, they used a modified ``24th-roots program" which lists all orthonormal bases whose vectors have 24th roots of unity as well as the number $b_2=(-1+2i)/{\sqrt{5}}$ (and its complex conjugate) as components.

In the numerical simulations realized with the physical imposition operator we have considered $10^4$ random choices of the parameter $c$. We found ten triplets of MU bases containing the pair $\{\mathbb{I},D(0)\}$, but only two of them are inequivalent, in agreement with \cite{Bengtsson1}. It is not difficult to find the analytic form of the triples by an educated guess. The exact value of the number $b_2$ was found again by imposing the unbiasedness of a solution to the pair $\{\mathbb{I},D(0)\}$.

Explicitly, the inequivalent triplets we find are $\{\mathbb{I},D(0),H_m\}$, $m=1,2$, where
\begin{equation}
H_1=
\left( \begin{array}{cccccc}
1    &   1  &   1    &  1    & 1     & 1\\
-i   &  -i  &  \omega^2   & \omega^{10}& \omega^{10}& \omega^2\\
-ib_2& ib_2 & \omega^9    & \omega^{21}& \omega^9   & \omega^{21}\\
-i   & -i   & \omega^{10} & \omega^2   & \omega^2   & \omega^{10}\\
ib_2 & -ib_2& \omega^{13} & \omega^{17}& \omega^5   & \omega\\
ib_2 & -ib_2& \omega^5    & \omega     & \omega^{13}& \omega^{17}\\
 \end{array} \right),
\end{equation}
\begin{equation}
H_2=
\left( \begin{array}{cccccc}
1      &    1    &   1    &  1    & 1     & 1\\
ib^*_2 &  -ib^*_2& \omega^{19} & \omega^7   & \omega^{23}& \omega^{11}\\
-ib^*_2& ib^*_2  & \omega^{15} & \omega^3   & \omega^3   & \omega^{15}\\
ib^*_2 & -ib^*_2 & \omega^{11} & \omega^{23}& \omega^7   & \omega^{19}\\
i      &    i    & \omega^{22} & \omega^{22}& \omega^{14}& \omega^{14}\\
i      &    i    & \omega^{14} & \omega^{14}& \omega^{22}& \omega^{22}\\
 \end{array} \right),
\end{equation}
and $\omega=e^{2\pi i/24}$. The MU vectors are given by the columns of the matrices $H_1$ and $H_2$. Interestingly, both of them are equivalent to a member of the Fourier family. We have verified that these analytical expressions are indeed solutions of the problem.

The two inequivalent triplets were found among the first three triplets obtained numerically, and no other inequivalent triplet was found in the next 100 runs of our program. This represents strong numerical evidence that no more than two inequivalent triplets exist which contain the pair $\{\mathbb{I},D(0)\}$. Moreover, both triplets occur with nearly equal frequency: we found $\{\mathbb{I},D(0),H_1\}$ 48 times while $\{\mathbb{I},D(0), H_2\}$ occurred 52 times, an observation which can be explained if one assumes that  the basin of attraction of every MU vector has the same size. This apparent symmetry has been noticed so far in each imposition-operator search for MU bases, whatever the  dimension $d$ \cite{Goyeneche1,Goyeneche2}.

Triplets of MU bases containing the pair $\{\mathbb{I},D_6^{(1)}(c)\}$ have also been found for many other values of the parameter $c$; none of the resulting triplets seems to allow for even a single further MU vector.

\section{Karlson's non-affine families}\label{non_extensi}
The most interesting property of the method defined in Section \ref{PIO} is that its computational costs do not increase when we consider non-affine families. This advantage can be used to analyze the construction of triplets from a pair of the form $\{\mathbb{I},H\}$ where $H$ belongs to a non-affine family of complex Hadamard matrices. Little seems to be known about extending such triplets, so that the method presented here is the first efficient way to study them over the entire parameter range of the families. We have performed computations for pairs of the form $\{\mathbb{I},K_6^{(2)}\}$ and $\{\mathbb{I},K_6^{(3)}\}$, and it seems that they cannot be extended to four MU bases (up to a possible null measure set of the parameters). Moreover, many pairs can be extended to a triplet only for a non-trivial subset of parameters. Let us start with the family $K_6^{(2)}$.

\subsection{Karlsson's biparametric family}\label{Karlsson_Section}
Karlsson has found a two-parameter non-affine family of complex Hadamard matrices $K_6^{(2)}$ in dimension six \cite{Karlsson}, which contains the families $D_6^{(1)},M_6^{(1)}$ and two subfamilies of the Fourier family. The Di\c{t}\u{a} family $D_6^{(1)}(t)$ is equivalent to the four corners $K_6^{(2)}(\pm\pi/2,\pm\pi/2)$ whereas $K_6^{(2)}(x,0)\sim F_6^{(2)}(x,x)$ and $K_6^{(2)}(0,x)\sim (F_6^{(2)}(x,x))^t$. Also, Matolcsi family determines one of the diagonals, that is, $K_6^{(2)}(x,x)\sim M_6^{(1)}(x)$. All these subfamilies are explicitly obtained from $K_6^{(2)}$ in Karlsson's paper \cite{Karlsson}. Note that a subset of the Fourier family and its transpose define the horizontal and vertical axes, respectively, of the parameter space of $K_6^{(2)}$. Also, the Fourier matrix $F_6^{(2)}(0,0)$ is equivalent to the center $K_6^{(2)}(0,0)$.

\begin{figure}[!h]
\centering 
{\includegraphics{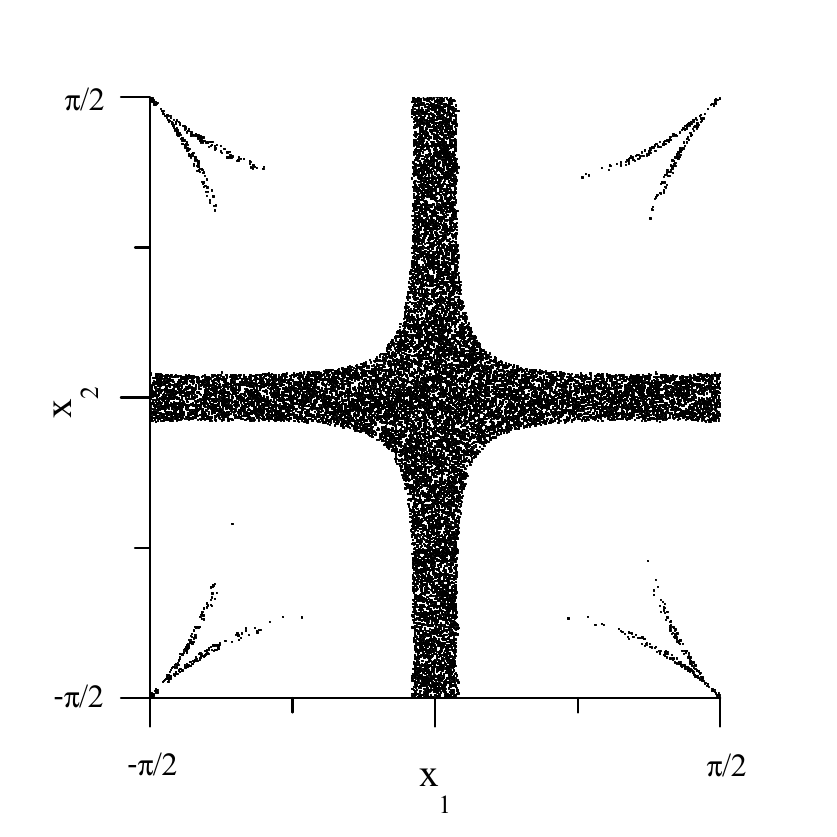}} 
\caption{Triplets of MU bases from $K_6^{(2)}$. A dot at the point $(x_1,x_2)$ indicates that the pair $\{\mathbb{I},K_6^{(2)}(x_1,x_2)\}$ can be extended to a triplet of MU bases.}
\label{Fig1}
\end{figure}

The results of our attempts to extend pairs of the form $\{\mathbb{I},K_6^{(2)}(x_1,x_2)\}$ to triplets are presented in Fig.\ref{Fig1}. In this figure, a black dot at the point $(x_1,x_2)$ means that a triplet has been found. Previously known result about affine families indicates that a triplet of MU bases can be obtained in the full range of the family \cite{Bengtsson1,Jaming2,Durt,Szollosi}, as far as we know. The evidence presented in Fig. \ref{Fig1} shows that triplets of MU bases only exist for a subset of parameters if the family is non-affine. In these simulations we have considered convergence of the sequences according to the upper bound 0.01 established in Eq.(\ref{bound}) and we have also considered the bounds 0.03, 0.05 and 0.08. In all the cases we found the same results, which evidence the stability of the solutions.

Furthermore, Fig. \ref{Fig1} clearly suggests the existence of new symmetries. The Fourier matrix $F_6$ and the Di\c{t}\u{a} family $D_6^{(1)}(t)$ seem to be privileged in the problem of constructing triplets from a pair of the form $\{\mathbb{I},K_6^{(2)}(x_1,x_2)\}$: both of them are centers of symmetries in Fig. \ref{Fig1}. Let us prove all symmetries existing in the family $K_6^{(2)}$, defined as:
\begin{equation}
K_6^{(2)}(x_1,x_2)=
\left( \begin{array}{cccccc}
1 &   1  &  1    &  1    & 1     & 1\\
1 &  -1  & z_1 & -z_1 & z_1 & -z_1 \\
1 & z_2  & -f_1 & -z_2f_2 & -{f_3}^* & -z_2{f_4}^*\\
1 & -z_2 & -z_1{f_2}^* & z_1z_2{f_1}^* & -z_1f_4 & z_1z_2f_3\\
1 & z_2 & -{f_3}^* & -z_2{f_4}^* & -f_1  & -z_2f_2\\
1 & -z_2 & -z_1f_4 & z_1z_2f_3 & -z_1{f_2}^* & z_1z_2{f_1}^*\\
 \end{array} \right),
\end{equation}
where $z_1=e^{ix_1}$ and $z_2=e^{ix_2}$, $-\pi/2\leq x_1,x_2\leq\pi/2$ and the four functions
\begin{eqnarray}
f_1&=&f(+x_1,+x_2),\nonumber\\
f_2&=&f(+x_1,-x_2),\nonumber\\
f_3&=&f(-x_1,-x_2),\nonumber\\
f_4&=&f(-x_1,+x_2),\label{efes}
\end{eqnarray}
are defined in terms of a single function, namely
\begin{equation}\label{generator}
f(x_1,x_2)=e^{i(x_1+x_2)/2}\left(\cos\left(\frac{x_1-x_2}{2}\right)-i\sin\left(\frac{x_1+x_2}{2}\right)\right)\left(\frac{1}{2}+i\sqrt{\frac{1}{1+\sin(x_1)\sin(x_2)}-\frac{1}{4}}\right).
\end{equation}
Karlsson has shown that
\begin{eqnarray}
f(x_1+\pi,x_2)&=&z_2f(x_1,-x_2),\\
f(x_1,x_2+\pi)&=&z_1f(-x_1,x_2),
\end{eqnarray}
meaning that
\begin{equation}
K_6^{(2)}(x_1+\pi,x_2)=K_6^{(2)}(x_1,x_2)P_{34}P_{56},
\end{equation}
and
\begin{equation}
K_6^{(2)}(x_1,x_2+\pi)=P_{36}P_{45}K_6^{(2)}(x_1,x_2),
\end{equation}
where $P_{34}$ and $P_{56}$ are permutations matrices. Consequently, one may restrict both parameters $x_1$ and $x_2$ to the interval $[-\pi/2,\pi/2]$.

\begin{figure}[!h]
\centering 
{\includegraphics{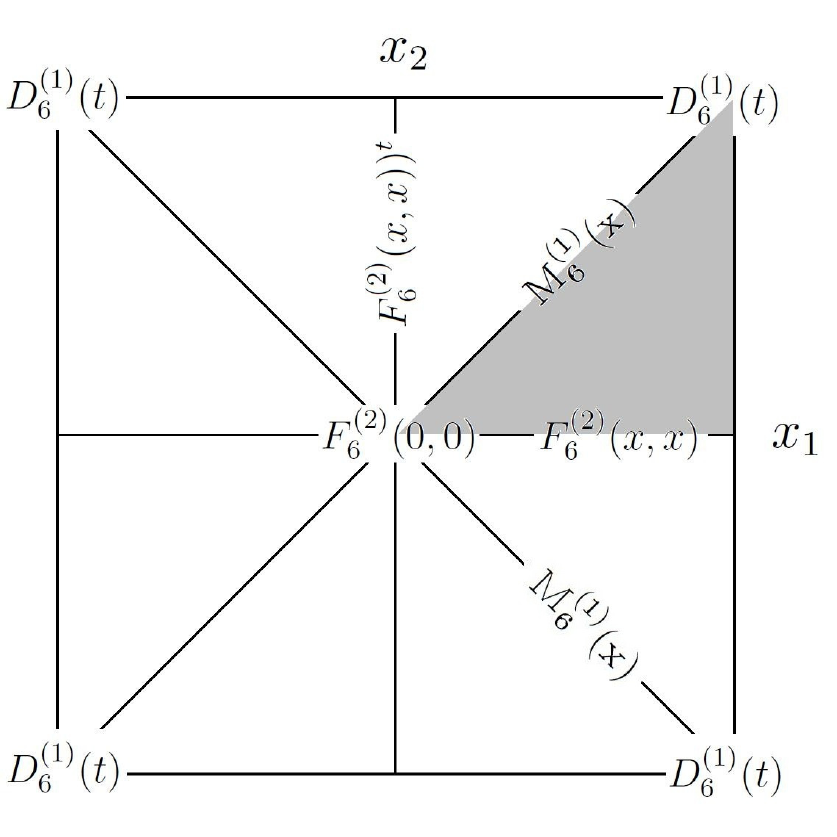}} 
\caption{The Karlsson's family $K_6^{(2)}(x_1,x_2)$ is equivalent to the gray triangle; note that $-\frac{\pi}{2}\leq x_j\leq\frac{\pi}{2},\,j=1,2$.}
\label{Fig2}
\end{figure}

Inspired by the symmetries of the graph shown in Fig. \ref{Fig1} and taking into account Eqs.(\ref{efes}) and Eq.(\ref{generator}), we notice that
\begin{eqnarray}
K_6^{(2)}(x_1,-x_2)&=&P_{36}P_{45}K_6^{(2)}(x_1,x_2),\label{K1}\\
K_6^{(2)}(-x_1,x_2)&=&P_{36}P_{45}K_6^{(2)}(x_1,x_2);\label{K2}
\end{eqnarray}
using the symmetry $f(x_1,x_2)=f(x_2,x_1)$, we also obtain
\begin{equation}
K_6^{(2)}(x_1,x_2)=K_6^{(2)}(x_2,x_1)\label{K3}.
\end{equation}
Eqs.(\ref{K1}) to (\ref{K3}) reveal the symmetry apparent in Fig. \ref{Fig1}, and we consider it unlikely that any further symmetries exist. The family $K_6^{(2)}(x_1,x_2)$ with $-\pi/2\leq x_1,x_2\leq\pi/2$ is divided into eight triangles of the same area, each of them containing one copy of the complete family. Therefore, it is sufficient to consider values in the triangle $x_1\in[0,\pi/2]$, $x_2\leq x_1$, i.e. the shaded area in Fig. \ref{Fig2}. In this figure, we show that the Matolcsi's family $M_6^{(1)}$ is located on the both diagonals of the square. As we will show later, we can construct triplets of MU bases from $M_6^{(1)}$, which means that it should appear in Fig. \ref{Fig1}. However, the set $\{M_6^{(1)}{x}\}$ is of measure zero within the family $K_6^{(2)}(x_1,x_2)$; since the parameters $(x_1,x_2)$ are chosen at random, the probability of observe it vanishes.

\subsection{Karlsson's tri-parametric family}
A tri-parametric non-affine family of complex Hadamard matrices has been recently found by Karlsson \cite{Karlsson3}, reading explicitly:

\begin{equation}
    K_6^{(3)}(\theta,\phi,\psi)=\left( \begin{array}{ccc}
F_2 &   Z_1  &  Z_2  \\
Z_3 &  \frac{1}{2}Z_3AZ_1  & \frac{1}{2}Z_3BZ_2  \\
Z_4 & \frac{1}{2}Z_4BZ_1  & \frac{1}{2}Z_4AZ_2
 \end{array} \right),
\end{equation}
where
\begin{equation}
F_2=\left( \begin{array}{cc}
1 &  1 \\
1 &  -1
\end{array} \right),\hspace{0.5cm}
A=\left( \begin{array}{cc}
A_{11} &  A_{12} \\
A_{12}^* &  -A_{11}^*
\end{array} \right),\hspace{0.5cm}
B=-F_2-A,
\end{equation}
and
\begin{equation}
Z_i=\left( \begin{array}{cc}
1 &  1 \\
z_i &  -z_i
\end{array} \right),\,i=1,2,\hspace{0.5cm}
Z_i=\left( \begin{array}{cc}
1 &  z_i \\
1 &  -z_i
\end{array} \right),\,i=3,4.
\end{equation}
The entries of $A$ are given by
\begin{eqnarray}
    A_{11}&=&-\frac{1}{2}+i\frac{\sqrt{3}}{2}(\cos(\theta)+e^{-i\phi}\sin(\theta)),\label{A11}\\
    A_{12}&=&-\frac{1}{2}+i\frac{\sqrt{3}}{2}(-\cos(\theta)+e^{i\phi}\sin(\theta)),\label{A12}
\end{eqnarray}
and the entries of $Z_i$ are
\begin{eqnarray}
z_1&=&e^{i\psi},\\
z_2^2&=&\mathcal{M}_A^{-1}(\mathcal{M}_B(z_1^2)),\label{Mobius1}\\
z_3^2&=&\mathcal{M}_A(z_1^2),\label{Mobius2}\\
z_4^2&=&\mathcal{M}_B(z_1^2).\label{Mobius3}
\end{eqnarray}
Here, $\mathcal{M}$ denotes the M\"{o}bius transformation, defined by
\begin{equation}
    \mathcal{M}(z)=\frac{\alpha z-\beta}{\beta^* z-\alpha^*},
\end{equation}
with $\alpha_A=A_{12}^2$, $\beta_A=A_{11}^2$, and $\alpha_B=B_{12}^2$, $\beta_B=B_{11}^2$, and $\theta,\phi,\psi\in[0,\pi)$.

This family contains the non-affine family $K_6^{(2)}$ and it also contains the complete set of the so-called $H_2$-reducible matrices. In dimension six, a complex Hadamard matrix is $H_2$-reducible if it contains nine $2\times2$ submatrices that are Hadamard matrices. Let us analyze an interesting particular case. It follows from Eq.(\ref{A11}) and (\ref{A12}) that the subfamily $K_6^{(3)}(0,\phi,\psi)$ do not depend on the parameter $\phi$. In this case, the M\"{o}bius transformations in Eqs.(\ref{Mobius1}--\ref{Mobius3}) turn into the identity irrespective of the value of $z$. Therefore, we obtain an affine one-parameter family
\begin{equation}
   K_6^{(3)}(0,\phi,\psi)=(P_{46} F_2\otimes F_3)\circ\mathrm{Exp}(iR(\psi)),
\end{equation}
which is contained in the Fourier family $F_6^{(2)}$. Here, the matrix $R(\psi)$ is defined by
\begin{figure}[h!]
\begin{center}
\subfigure[\label{fig3_1}\hspace{0.2cm}Triplets for $K_6^{(3)}(\theta,\phi,0)$]{
\includegraphics{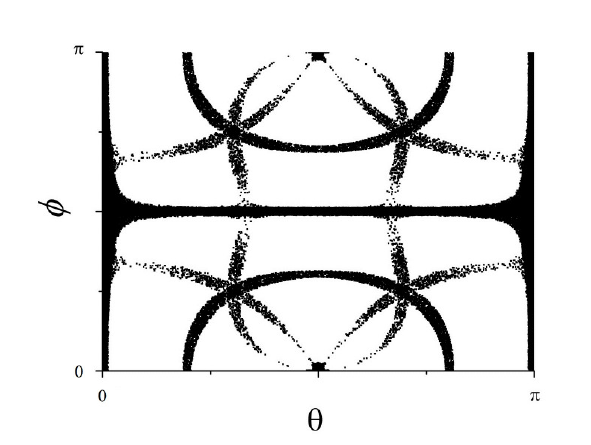}}
\subfigure[\label{fig3_2}\hspace{0.2cm}Triplets for $K_6^{(3)}(\theta,\phi,\pi/4)$]{
\includegraphics{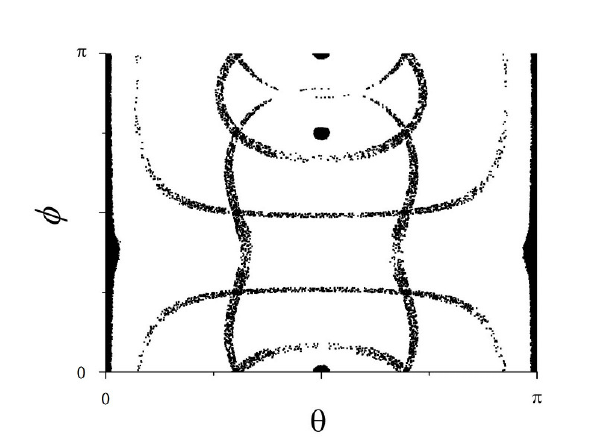}}
\subfigure[\label{fig3_3}\hspace{0.2cm}Triplets for $K_6^{(3)}(\theta,\phi,\pi/2)$]{
\includegraphics{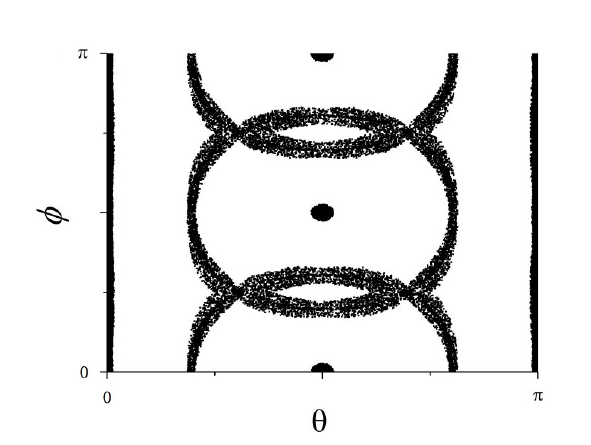}}
\subfigure[\label{fig3_4}\hspace{0.2cm}Triplets for $K_6^{(3)}(\theta,\phi,3\pi/4)$]{
\includegraphics{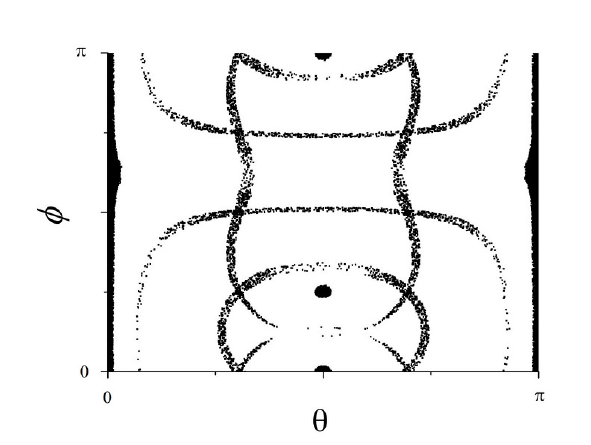}}
\caption{Triplets in Karlsson's family $K_6^{(3)}(\theta,\phi,\psi)$ for some fixed values of $\psi$.}
\end{center}
\end{figure}
\begin{equation}
    R(\psi)=
\left( \begin{array}{cccccc}
\bullet\hspace{0.2cm}&\bullet\hspace{0.2cm}&\bullet\hspace{0.2cm}&\bullet\hspace{0.2cm}&\bullet\hspace{0.2cm}&\bullet\\
\bullet&\bullet&\psi&\psi&\bullet&\bullet\\
\bullet&\bullet&\bullet&\bullet&\bullet&\bullet\\
\bullet&\bullet&\psi&\psi&\bullet&\bullet\\
\bullet&\bullet&\bullet&\bullet&\bullet&\bullet\\
\bullet&\bullet&\psi&\psi&\bullet&\bullet
\end{array} \right),
\end{equation}
where $\bullet$ means a null entry. The permutation matrix $P_{46}$ interchanges rows 4 and 6. This subfamily allows a triplet for any $\psi\in[0,\pi)$, because it is contained in the Fourier family $F_6^{(2)}(a,b)$, which admit a triplet for any value of $a$ and $b$.

Numerical simulations from considering the family $K_6^{(3)}$ are shown in Fig. \ref{fig3_1} to Fig. \ref{fig3_4}. As we can see, these figures strongly suggest the existence of reflection symmetries in the three variables $\theta,\phi$ and $\psi$. However, we have not been able to find them analytically even in the simplest case $\psi=0$. There are highly non-trivial diagonal matrices and row permutations which do not allow us to reveal the hidden symmetries. On the other hand, if a random value of $\psi$ is considered we always obtain the same kind of two dimensional objects. No fractal structures have been detected for any value of the parameters. Consequently, we have a representative description of the general case, in the sense that the evolution of the parameter $\psi$ gives us a smooth connection between these four figures (Fig. \ref{fig3_1} to Fig. \ref{fig3_4}).

\section{Summary and conclusions}\label{Summary}
We have presented an efficient numerical method to construct sets of mutually unbiased bases in finite dimension. The main advantage of our method appears when non-affine families of complex Hadamard matrices are considered, where the standard method to solve coupled polynomial equations (Buchberger's algorithm) often stalls due to excessive memory requirements. Our method numerically solves the problem for non-affine families with the same computational cost as for affine families.

To test our approach we first used it to determine the well known 48 MU vectors to the pair $\{\mathbb{I},F_6\}$. We have been able to prove that they determine three orbits under the Weyl-Heisenberg group. Also, considering the Di\c{t}\u{a} matrix we found two triplets that agree with a result obtained previously \cite{Bengtsson1}.

\begin{table}[htdp]
\begin{center}
\begin{tabular}{|c|c|c|}
\hline
Pair & Kind & Maximal set of MU bases \\
\hline
$\{\mathbb{I},S_6^{(0)}\}$ & \hspace{0.3cm} Isolated \hspace{0.3cm} & 2 \\
\hline
$\{\mathbb{I},D_6^{(1)}(c)\}$ & Affine & $3,\, \forall c\in[-\frac{1}{8},\frac{1}{8}]$ \\
\hline
$\{\mathbb{I},F_6^{(2)}(a,b)\}$ & Affine & $3,\,\forall a,b\in[0,2\pi)$ \\
\hline\hline
$\{\mathbb{I},B_6^{(1)}(s)\}$ & Non-affine & $3,\, \forall s\in[-\pi,\arccos(\frac{-1+\sqrt3}{2})]\cup[\arccos(\frac{-1+\sqrt3}{2}),-\pi]$ \\
\hline
$\{\mathbb{I},M_6^{(1)}(t)\}$ & Non-affine & $\left\{
\begin{array}{c l}
 3 & \mbox{if } t\in[0.5309\pi,0.9157\pi]\\
 3 & \mbox{if }t\in[1.5312\pi,1.9163\pi]\\
 2 & \mbox{otherwise}
\end{array}
\right.
$ \\
\hline
$\{\mathbb{I},K_6^{(2)}(x_1,x_2)\}$ & Non-affine & $\left\{
\begin{array}{c l}
 3 & \mbox{in black regions of Fig. \ref{Fig1}}\\
 2 & \mbox{in white regions of Fig. \ref{Fig1}}
\end{array}
\right.
$\\
\hline
$\{\mathbb{I},K_6^{(3)}(\theta,\phi,\psi)\}$ & Non-affine & $\left\{
\begin{array}{c l}
 3 & \mbox{in black regions from Fig. \ref{fig3_1} to \ref{fig3_4}}\\
 2 & \mbox{in white regions from Fig. \ref{fig3_1} to\ref{fig3_4}}
\end{array}
\right.
$\\
\hline
\end{tabular}
\end{center}
\caption{Maximal set of MU bases for some families. The last four cases are new results.}
\label{tabla}
\end{table}

Table \ref{tabla} summarizes the results obtained in this paper, indicating the maximal set of MU bases that can be constructed from pairs of MU bases associated with various families in dimension six. In the cases of affine families and the non-affine family $B_6^{(1)}$ we found triplets in the entire range of the parameters, whereas the isolated matrix $S_6^{(0)}$ does not allow a triplet. This property of $S_6^{(0)}$ has been previously found by Brierley and Weigert \cite{Brierley6}.

In the simulation realized for the non-affine family $B_6^{(1)}$ we have considered 10.000 random choices of its parameter, and we obtained a triplet in every case. This is the only non-affine family where we found a triplet for every value. The non-affine family $M_6^{(1)}$ defined in the range $t\in(\pi/2,\pi]\cup(3\pi/2,2\pi]$ does not allow a triplet in its full range. We have realized three simulations considering 1,000; 10,000 and 100,000 random choices of the parameter $t$ and we obtain the same results. That is, a part of the family $M_6^{(1)}$ does not allow us to construct a triplet of MU bases (see Table \ref{tabla}).

In the cases of Karlsson's families $K_6^{(2)}$ and $K_6^{(3)}$ we have considered 2 millon and 8 millon random choices, respectively, sampling the entire parameter ranges of both families. We have shown that these two families extend to triplets at most. The property that a triplet of MU bases can be found only for a reduced set of parameters of a family is a new result presented here by the first time. In addition, we identified new symmetries that reduce the range of the parameters of the family $K_6^{(2)}$.

Finally, in our investigation of more than ten million complex Hadamard matrices belonging to non-affine families we could not find a \emph{single vector being MU} to a triplet. This evidence supports the conjecture that no more than three MU bases can be constructed in dimension six.
\section{Acknowledgments}
I specially thank to Stefan Weigert for his invaluable help in order to make possible this article. This work is supported by Grants FONDECyT N$^{\text{\underline{o}}}$ 3120066 and MSI P010-30F.


\begin{thebibliography}{99}
\bibitem{Bennett} C.H. Bennett and G. Brassard. Quantum cryptography: Public key distribution and coin tossing. Proceedings of the IEEE International Conference on Computers, Systems and Signal Processing, 1984.
\bibitem {Brub} D. Bru\ss. Optimal eavesdropping in quantum cryptography with six states. PRL, 81:3018-3021 (1998).
\bibitem {Cerf} N. J. Cerf, M. Bourennane, A. Karlsson, and N. Gisin. Security of quantum key distribution using d-level systems. Phys. Rev. Lett., 88(12):127902 (2002)
\bibitem {Spengler} C. Spengler et al. Entanglement detection via mutually unbiased bases. Phys. Rev. A 86, 022311 (2012).
\bibitem{Durt} T. Durt, B. Englert, I. Bengtsson, K. \.{Z}yczkowski. On Mutually unbiased bases. Int. Jour. of Quant. Inf. \textbf{8}, 4, 535-640 (2010).
\bibitem {Gibbons} K.S. Gibbons, M.J. Hoffman, and W.K. Wootters. Discrete phase space based on finite fields. Phys. Rev. A, 70(062101):1-23, 2004.
\bibitem{Bengtsson3} Ingemar Bengtsson. Mubs, polytopes, and finite geometries. AIP conference Proceedings, 750:63-69, 2005.
\bibitem{Aharonov} Y. Aharonov and B. G. Englert, Z. Naturforsch. A: Phys. Sci. 56a, 16 (2001).
\bibitem{Zauner} G. Zauner. Ph.D. Thesis, University of Wien (1999).
\bibitem{Archer} C. Archer. There is no generalization of known formulas for mutually unbiased bases. J. Math. Phys. \textbf{46}, 022106 (2005).
\bibitem{Jaming} P. Jaming, M. Matolcsi, P. M\'{o}ra. The problem of mutually unbiased bases in dimension 6. Cryptography and Communications \textbf{2}, 211-220 (2010).
\bibitem{Grassl} M. Grassl. On SIC-POVMs and MUBs in Dimension 6, in: Proc. ERATO Conference on Quantum Information Science (EQUIS 2004).
\bibitem{Butterley} P. Butterley, W. Hall. Numerical evidence for the maximum number of mutually unbiased bases in dimension six- Phys. Lett. A \textbf{369}, 5-8 (2007).
\bibitem{Bengtsson1} I. Bengtsson, W. Bruzda, A. Ericsson, J. Larsson, W. Tadej and K. \.{Z}yczkowski. MUBs and Hadamards of order six. J. Math. Phys. \textbf{48}, 052106 (2007).
\bibitem{Brierley} S. Brierley, S. Weigert. Maximal Sets of Mutually Unbiased Quantum States in Dimension Six. Phys. Rev. A \textbf{78}, 042312 (2008).
\bibitem{Brierley3} S. Brierley, S. Weigert. Mutually Unbiased Bases and Semi-definite Programming, J. Phys.: Conf. Ser. \textbf{254}, 012008 (2010).
\bibitem{Jaming2} Jaming P. et al., A generalized Pauli problem and an infinite family of MUB-triplets in dimension 6, J. Phys. A: Math. Theor. \textbf{42}, 245305 (2009).
\bibitem{Brierley4} S. Brierley. Mutually Unbiased Bases in Low Dimensions. PhD thesis, University of York (2009).
\bibitem{Ivanovic} I. Ivanovic, J. Phys. A \textbf{14}, 3241. (1981).
\bibitem{Wootters} W. Wootters and B.  Fields, Ann. Phys. \textbf{191}, 363 (1989).
\bibitem{Bengtsson2} Bengtsson I., Three ways to look at mutually unbiased bases. Found. Prob $\&$ Phys. 4 AIP Conf. Proc. \textbf{889}, 40-51 (2007). Preprint: quant-ph/0610216 (2006).
\bibitem{Bruzda}
\url{http://chaos.if.uj.edu.pl/~karol/hadamard}
\bibitem{Brierley2} S. Brierley, S. Weigert, O. Bengtsson. All mutually unbiased bases in dimensions two to five. Int. J. Quant. Info. \textbf{10} 803 (2009).
\bibitem{Buchberger} B. Buchberger, An Algorithm for Finding the Basis Elements of the Residue Class Ring of a Zero Dimensional Polynomial Ideal. Ph.D. Dissertation, University of Innsbruck (1965) (English translation by M. Abramson in J. Symb. Comp. \textbf{41}, 471 (2006)).
\bibitem{McNulty} D. McNulty, S. Weigert. On the Impossibility to Extend Triples of Mutually Unbiased Product Bases in Dimension Six. Int. J. of Quant. Inf. \textbf{10}, 5 (2012).
\bibitem{Goyeneche1} D. M. Goyeneche, A. C. de la Torre. State determination: An iterative algorithm. Phys. Rev. A \textbf{77}, 042116 (2008).
\bibitem{Goyeneche2} D. Goyeneche, A. de la Torre. Quantum state reconstruction from dynamical systems theory. arXiv:1103.3213 (2011).
\bibitem{Hellinger} E. Hellinger. Die Orthogonalinvarianten quadratischer Formen von unendlich vielen Variablen. Dissertation, Göttingen (1907).
\bibitem{Brierley6} S. Brierley, S. Weigert. Constructing mutually unbiased bases in dimension six. Phys. Rev. A \textbf{79}, 052316 (2009).
\bibitem{Bjorck1} G. Bj\"{o}rck and R. Fr\"{o}berg, A faster way to count the solutions of inhomogeneous systems of algebraic equations, with applications to cyclic n-roots, J. Symbolic Computation \textbf{12}, 329 (1991).
\bibitem{Bjorck2} G. Bj\"orck and B. Saffari, New classes of finite unimodular sequences with unimodular Fourier transforms. Circulant Hadamard matrices with complex entries, C. R. Acad. Sci. Paris, Ser. I \textbf{320}, 319 (1995).
\bibitem{McNulty2} D. McNulty, S. Weigert. J. Phys. A: Math. Theor. \textbf{45} 135307 (2012).
\bibitem{Goyeneche3} A. de la Torre, D. Goyeneche. Quantum mechanics in finite-dimensional Hilbert space. Am.J.Physics \textbf{71}, 1, 49-54 (2003).
\bibitem{Moroz} B.Z. Moroz, A.M. Perelomov. On a problem posed by Pauli. Theor. Math. Phys. \textbf{101}, 1200-1204 (1994).
\bibitem{Bandyo} S. Bandyopadhyay \emph{et Al.} A new proof of the existence of mutually unbiased bases. Algorithmica \textbf{34}, 512-528 (2002).
\bibitem{Dita} P. Di\c{t}\u{a}, Some results on the parametrization of complex Hadamard matrices, J. Phys. A: Math. Gen. \textbf{37}, 5355 (2004).
\bibitem{Karlsson} B. Karlsson. Two-parameter complex Hadamard matrices for $N=6$. J. Math. Phys. \textbf{50}, 082104 (2009)
\bibitem{Szollosi} F. Sz\"oll\H{o}si. Construction, classification and parametrization of complex Hadamard matrices (PhD thesis) arXiv:1110.5590 (2011).
\bibitem{Karlsson3} B. Karlsson. Three-parameter complex Hadamard matrices of order 6. Lin. Alg. Appl. \textbf{434}, 247 (2011).
\end{thebibliography}
\end{document}